\documentclass[12pt]{article}
\usepackage{graphicx}
\usepackage{epstopdf}
\def\singlespace {\smallskipamount=3.75pt plus1pt minus1pt
                  \medskipamount=7.5pt plus2pt minus2pt
                  \bigskipamount=15pt plus4pt minus4pt
                  \normalbaselineskip=15pt plus0pt minus0pt
                  \normallineskip=1pt
                  \normallineskiplimit=0pt
                  \jot=3.75pt
                  {\def\smallskip {\vskip\smallskipamount}}
                  {\def\medskip   {\vskip\medskipamount}}
                  {\def\bigskip   {\vskip\bigskipamount}}
                  {\setbox\strutbox=\hbox{\vrule
                    height10.5pt depth4.5pt width 0pt}}
                  \parskip 7.5pt
                  \normalbaselines}
\def\middlespace {\smallskipamount=5.825pt plus1.5pt minus1.5pt
                  \medskipamount=11.25pt plus3pt minus3pt
                  \bigskipamount=22.5pt plus6pt minus6pt
                  \normalbaselineskip=22.5pt plus0pt minus0pt
                  \normallineskip=1pt
                  \normallineskiplimit=0pt
                  \jot=5.825pt
                  {\def\smallskip {\vskip\smallskipamount}}
                  {\def\medskip   {\vskip\medskipamount}}
                  {\def\bigskip   {\vskip\bigskipamount}}
                  {\setbox\strutbox=\hbox{\vrule
                    height15.75pt depth6.75pt width 0pt}}
                  \parskip 7.25pt
                  \normalbaselines}
\def\dblspc {\smallskipamount=7.5pt plus2pt minus2pt
                  \medskipamount=15pt plus4pt minus4pt
                  \bigskipamount=30pt plus8pt minus8pt
                  \normalbaselineskip=30pt plus0pt minus0pt
                  \normallineskip=2pt
                  \normallineskiplimit=0pt
                  \jot=7.5pt
                  {\def\smallskip {\vskip\smallskipamount}}
                  {\def\medskip   {\vskip\medskipamount}}
                  {\def\bigskip   {\vskip\bigskipamount}}
                  {\setbox\strutbox=\hbox{\vrule
                    height21.0pt depth9.0pt width 0pt}}
                  \parskip 15.0pt
                  \normalbaselines}

\def\be{\begin{equation}}

\def\j-{\J_-}

\def\ee{\end{equation}}

\def\bearr{\begin{eqnarray}}
\def\bearrs{\begin{eqnarray*}}
\def\eearr{\end{eqnarray}}
\def\eearrs{\end{eqnarray*}}
\def\barr{\begin{array}}
\def\earr{\end{array}}

\def\nn8{\nonumber\\[10pt]}

\def\dis{\displaystyle}

\def\ed{\end{document}}

\begin{document}
\singlespace


\begin{center}
{\bf Beta decay rates for r-process for nuclei near neutron number $N=82$} 
\vskip 0.25cm

Kamales Kar$^{a,}$\footnote{ Email: kamales.kar@saha.ac.in Tel: (91) 33 2337 5345 FAX: (91) 33 2337 4637} and Soumya Chakravarti$^{b,}$\footnote{Email: chakra@csupomona.edu} \\ \  \\
$^{a}${\it Saha Institute of Nuclear Physics, 1/AF
Bidhan Nagar, Calcutta 700 064, India} \\
$^{b}${\it Department of Physics, California State Polytechnic University, Pomona, CA~91768, USA}\\
\end{center}
\vskip 0.5cm

{\small
\noindent {\bf ABSTRACT:} For r-process nucleosynthesis the beta decay rates
 of very neutron-rich nuclei are important ingredients. We consider the region
around the neutron number N=82 and calculate the half-lives and rates for a
number of nuclei. Forms for beta strength functions based on spectral distribution
methods are used. The calculated half-lives are first compared to the observed
values and then predictions are made for very neutron-rich nuclei close to
drip line for which no experimental values are available.  

\bigskip



\bigskip

\begin{center}
{\bf 1. Introduction}
\end{center}

The production of elements heavier than iron takes place through the s-process
and r-process nucleosynthesis \cite{burb-57}. Though roughly half the nuclei
with $A>56$ are created through the r-process, its sites are still uncertain.
Many candidates have been proposed and the most favored ones are all connected
to core collapse supernovae. But they have the drawback that the mechanism of
explosion of these supernovae is not completely understood yet and almost all
simulations with realistic physics input fail to reproduce the explosion
\cite{arnould-07}. The neutrino driven wind outside a protoneutron star
for a delayed core-collapse supernova (DCCSN) 
is considered to 
be a promising site. In the environments of high temperatures 
 (a few times $10^{9}$ K) and high neutron density ($>10^{20} \rm cm^{-3}$)
and high entropy, neutron capture is rapid compared to the
lifetime of beta decay. So normally a succession of neutron captures takes place
followed by a waiting point and then a beta decay. The waiting point is dominant
near shell closures. So through  r-process nucleosynthesis often very 
neutron-rich nuclei far away from the valley of stability are produced. As the
r-process path goes through the magic nuclei with  N=50, 82 and 126, these nuclei can
clearly be identified by the observed nuclear abundance peaks \cite{burb-57}. 
Because of the uncertainties of the sites, a number of site free parametric
high temperature r-process models proposed over the years, like the canonical
r-process model (CAR) \cite{Seeger-65}, the multi-event r-process model (MER)
\cite{Goriely-96} \cite{Bouquelle-95} etc., are quite useful. 
 Over the last two decades radioactive ion beams have been extensively
used to create many of the nuclei in the possible r-process path. 
But some of the nuclei near the drip line are so neutron rich it is difficult
to make any detailed measurement in the laboratory. So the
knowledge of the nuclear properties of these very large number of nuclei is 
still far from complete. The nuclear physics of the r-process for these nuclei 
needs information on  ground state properties, $(n,\gamma)$ and $(\gamma,n)$ rates,
$\beta^{-}$ half-lives, beta-delayed neutron emission and different modes of
fission. In the absence of experimental data results of theoretical calculations
are very useful. Though one prefers microscopic models one expects them to
be universal. So theoretical models applicable globally or  at least over a
region of nuclear masses are of particular relevance. Takahashi and Yokoi
\cite{Takahashi-87} calculated beta decay rates for heavy nuclides ($25<Z<84$
and $58<A<210$) at temperatures $5\times 10^{3} <T<5\times 10^{7}$ K and electron
density $10^{20} < n_{e} < 3\times 10^{27} \rm cm^{-3}$. Though this is for all nuclei
going through beta decay and electron capture in the s-process paths 
 the method is very general and has wider applicability. They
estimate unknown $ft$-values from sytematics.

Calculation of beta decay and electron capture rates has been pursued 
vigorously during the last twenty years for pre-supernova evolution and
gravitational collapse stage for core collapse supernovae \cite{lmp-03}.
Whereas for the sd-shell nuclei the rates are given by Oda {\it et al.}  \cite{Oda-94},
for the fp-shell nuclei the works of Fuller, Fowler and
Newman \cite{ffn-80} \cite{ffn-82a} \cite{ffn-82b} were used as the standard reference  for quite some time. Aufderheide 
\cite{auf-91} \cite{auf-94} indicated the usefulness of interacting shell model for the
rates. They also calculated approximate abundances in stellar conditions 
appropriate for the pre-supernova and collapse stage  and identified the nuclei which are the most important in
the network for each density-temperature grid point. On the other hand
statistical methods for the beta decay strength  distribution were used to
calculate the rates for fp-shell nuclei \cite{kar-94}. This was followed by large
space and detailed shell model calculations of the allowed beta decay strength
to calculate  weak interaction rates for nuclei with $A<65$ \cite{lmp-03}
\cite{lmp-01}. The effect of the new shell model rates on the late stage of
evolution of massive stars has also been seen \cite{heger-01}. Recently 
Pruet and Fuller \cite{pruet-03} used FFN-type ideas supplemented by 
experimental information wherever available to calculate the rates for nuclei
in the mass range $A=65-80$ relevant for supernova collapse. 

 There are two classes of models for calculation of beta decay half-lives and
rates for neutron rich nuclei. Among the macroscopic models the Gross Theory of
beta decay \cite{Yamada-65} \cite{Takahashi-69} \cite{Yako-05} \cite{Takahashi-71}
is one of the earliest models that took into account the Gamow-Teller (GT) giant
resonances in some form for nuclei far away from stability. The later
improvements \cite{Tachibana-90} \cite{Nakata-97} known as Semi-Gross Theory
or 2nd generation Gross Theory are used for r-process calculations \cite{arnould-07}. 
Among the microscopic approaches to estimate half-lives of a large set of nuclei
different versions of a combination of global mass models and Quasiparticle
Random Phase Approximation (QRPA) have been used.
 These are FRDM/QRPA \cite{moll-97} and
ETSI/QRPA \cite{bor-00}. Self-consistent Hartree-Fock-Bogoliubov plus
QRPA model has also been used \cite{engel-99}.

The shell closure at $N=82$ for n-rich nuclei in the r-process path is
receiving a lot of attention lately and more realistic models for beta decay
half-lives and rates in this region are needed now. We construct here a model
 for the beta decay strength function based on the experience with lighter nuclei in
the fp shell. This takes into account the tail of the GT giant resonance with
the centroid of the resonace empirically fixed and its width obtained by best
fit to the known neutron-rich nuclei.
We deal with nuclei for which a few lowlying log ft's are known,
to test the
model and then also apply  it to cases where not much is known experimentally
beyond the Q-value and half-life.  
We then use it to predict the experimentally unknown half-lives of nuclei near
the drip line where very little data are available. These nuclei may play 
important roles in the r-process. The actual beta decay rates are also 
calculated for stellar
density and temperatures which are thought to be typical for the r-process.
Section 2 describes the model that we construct to calculate the half-lives 
and rates and section 3 gives the results for such calculations for the
neutron-rich nuclei in the mass range of $115<A<140$. 

\newpage

\begin{center}
{\bf 2. Model For Beta Strength Function}
\end{center}

In stellar situations the beta decay of a nucleus, in principle, can take place
not only from the ground state of the mother but from the low-lying excited
states as well. The contributions from the excited states depend on the
temperature.
Thus the beta decay rate for a nucleus at temperaure T is
given by

\be
\barr{rcl}
\lambda= \dis\sum_{i}\; e^{-E_i/kT} \dis\sum_{j} \lambda_{ij} / Z
\earr
\ee

where $E_i$ is the energy of the state of the mother nucleus $|i>$ and Z is the
partition function of the nucleus. `$j$' sums over final states of the daughter 
nucleus $|j>$ to which transitions are possible and $\lambda_{ij}$
 stands for the rate between these two states. For allowed decay this rate 
 is given by

\be
\barr{rcl}
\lambda_{ij}= \frac{\displaystyle \ln\ 2}{\displaystyle (ft)_{ij}} f_{ij} =
 \ln\ 2\ f_{ij} \left[ \frac{\displaystyle |M_{F}|^{2}}{\displaystyle 10^{3.79}}
+ \frac{\displaystyle |M_{GT}|^{2}}{\displaystyle 10^{3.59}} \right]
\earr
\ee

In eq (2) the first part in the brackets is the contribution from the
 Fermi operator and the second part from the Gamow-Teller operator. 
Also $f_{ij}$ is the phase space factor for the beta decay with the
emitted electron with energy $E$ and rest mass energy $m_{e}c^{2}$ going to the top of the electron Fermi sea outside and has
the expression
\be
\barr{rcl}
f_{ij}= \int_{1}^{\epsilon_{0}} F(Z,\epsilon) (\epsilon_{0}-\epsilon)^{2} \epsilon (\epsilon^{2}-1)^{1/2} (1-f_{e})d\epsilon 
\earr
\ee
with $\epsilon_{0}=E_{ij}/m_{e}c^{2}$ and $\epsilon=E/m_{e}c^{2}$ where $E_{ij}$
is the maximum energy the electron can have for the transition from the state $|$i$>$ to the state $|$j$>$.
$f_{e}$ is the Fermi-Dirac distribution of the electrons outside the nucleus at temperature T.
For the antineutrinos under consideration, there is no buildup of antineutrino gas  outside.
Here $F(Z,\epsilon)$ is the Coulomb correction factor and we use the expression
of \cite{schent-83}. In eq (2) the Fermi and the Gamow-Teller matrix elements for $\beta^{-/+}$ decay are given by
\be
\barr{rcl}
|M_{F}(fi)|^{2}=  |<f| \Sigma _{k} t_{-/+}(k) |i> |^{2} /(2 J_{i} +1)
\earr
\ee
\be
\barr{rcl}
|M_{GT}(fi)|^{2}=  |<f| \Sigma _{k} {\vec {\sigma}}(k) t_{-/+}(k) |i> |^{2} /(2 J_{i} +1)
\earr
\ee

For calculation of half-lives the decay is from the ground state, {\it i.e.}, there is
only one mother state and the Boltzmann factor is equal to 1 in eq (1). 

The beta decay strength distribution has contributions from a number of discrete
low-lying states of the daughter and the GT giant resonance. However the Fermi
resonance is a very sharp one and centred around the Isobaric Analog State 
(IAS), which gets pushed up in energy above the ground state of the mother
by the Coulomb interaction. In our model 
 we take the resonance
width as $ \sigma_c = 0.157 Z A^{-1/3}$ \cite{kar-94} \cite{morita-73}. As
the spreading width is small this resonance cannot be reached by Q-value and
so Fermi makes very little contribution to half-lives and rates.  

However the GT giant resonance is a broad one due to the spin dependence of the
Hamiltonian and the GT operator and the tail of the giant resonance can be 
reached by the Q value. Spectral distribution theory suggests that for a
large number of valence nucleons and for very large shell model spaces, the
strength distribution in final energy for one-body and similar excitation
operators asymptotically goes towards a Gaussian \cite{French-88} \cite{kotakar-89}.
But the actual GT strength distribution for nuclei often deviate substantially
from Gaussian particularly in the tail region. However in this work we shall
take the form of the beta strength coming from the GT giant resonance to
be a Gaussian. For the low-lying states of the daughter whose $\log ft$'s
are known we explicitly take into account these strengths in the summation
of eq (1). Their strengths are subtracted from the GT sum rule strength used
for the giant resonance, {\it viz.}  $3(N-Z)$, where $N$ and $Z$ are the
neutron and proton numbers of the mother. This expression is known as the Ikeda 
sum expression \cite{Ikeda-63} and normally gives the difference of the 
$\beta^{-}$ and the $\beta^{+}$ strengths. But for our cases as the allowed
neutron orbitals for the valence protons are blocked
 the $\beta^{+}$ sum rule strength
is negligibly small. For the GT centroid, 
Bertsch and Esbensen \cite{bert-87} using the Tamm-Dancoff Approximation give the
following expression \cite{pruet-03}

\be
\barr{rcl}
     E_{GT^-} = E_{IAS}+ \Delta E_{s.o.} + 2 [ k_{\sigma \tau} S_{GT^-}/3 - (N-Z) k_{\tau} ]
\earr
\ee

where $\Delta E_{s.o.}$ is the contribution coming from the spin-orbit force.
and we use the orbit averaged value of 3.0 MeV \cite{bert-87} for it.
The last two terms have their origin in the spin-isospin and isospin dependent  
nuclear forces.  
Like Pruet and Fuller \cite{pruet-03} we use the values $k_\tau = 28.5/A$ and $k_{\sigma\tau} = 23/A$.
It is wellknown that the actual GT strength sums are quenched compared to the 
theoretical values. We take this into account by an overall quenching factor
of 0.6 in the sum rule \cite{kar-94}.
The width of the Gaussian is left as a free parameter to best reproduce the 
half-lives of decaying nuclei in the range $115<A<140$. The $\beta^{-}$ decays
with $Q$ values greater than 5 MeV are selected as one needs enough final states
to use the statistical averaged strength function. Decays with lower $Q$ values
are normally dominated by specific transitions to one or two states and for
them microscopic models are required. The nuclei in this range are divided
into two groups: i) nuclei for which some low-lying $\log ft$'s are known, and ii)
nuclei for which no measured low-lying $\log ft$ is available. The second group
 of nuclei are the more neutron rich ones and on the average have larger $Q$ 
values. As the r-process nuclei roughly have $Q$ values of 10 MeV or more,
again the second group are the ones that are more relevant for r-process.
The width extracted for the second group is the one that can be used for the
prediction of half-lives of nuclei near the drip-line taking part in the
r-process. This is done for a number of isotopes in this mass range for which
very little experimental information is available.

We also observe that for odd-$N$ nuclei with $N>82$, as the neutrons
and protons occupy different shells, the parity of the ground state of the mother
is different from the parity of the states in the ground state region of the 
daughter. So there are normally no allowed
transitions from the ground state of the mother to the low-lying states of
the daughter and instead one has much weaker transitions coming from the
forbidden beta decays.
The GT transitions
 take place to states at higher energies and our model can handle that 
situation too.

\begin{center}
{\bf 3. Results}
\end{center}

We first consider the set of nuclei in the mass range of 115 to 140 for which
some low-lying $\log ft$ values are known \cite{tabisotope-03}. The width of the
Gamow Teller resonance is varied to minimise the quantity $S= \Sigma_{n} 
 [ \log \left(\tau_{1/2,\rm calc}/\tau_{1/2,\rm expt}\right)]^2$
summed over all the n=32 nuclei. Table 1 gives the calculated values of the
half-lives and compares them with the experimental values \cite{tabisotope-03}.
The best fit value of the width is 4.2 MeV and for that width S/n turns out to
be 0.2630. One finds that 20 of the calculated values are within a factor of
2 from the experimental half-lives while 9 more are within a factor of 5. Only
the nuclei $^{119}$Ag, $^{122}$In and $^{123}$Ag have their calculated values
off by factors of ten or more and all three have the experimental values much 
smaller. Excluding them the value of $S/n$ turns out to be 0.0744. Excluding
only $^{122}$In and $^{123}$Ag makes it 0.1088.

Table 2 gives the calculated and experimental half-lives for 37 nuclei in the  
same mass range. The best fit value for the GT width is 6.9 MeV. The value for
S/n in this case is 0.1131. 23 of the calculated half-lives are within a factor
of 2 from the observed values and the remaining 14 are all within a factor 5.  
The value of the width turns out to be larger than the GT giant resonance width.
However we point out that in our model, the resonance has been artificially made wider to
take account of some of the discrete strengths to states lying below the
giant resonance.
Figure 1 shows the cases without known $\log ft$'s alongwith the cases 
with the known ones, i.e., 69 cases in all with the value of 
$\log\left({\tau_{1/2,\rm calc}/\tau_{1/2,\rm expt}}\right)$ plotted as a function of the $Q$ value. 
In Figure 2 we plot the calculated half-lives of all the 69 nuclei and compare them with
a simple form which depends only on the $Q$ value given by
\be
\barr{rcl}
\lambda_{\beta^{-}}= \ln \ 2 /t_{1/2}= 10^{-4} \times (Q_{\beta^{-}}/MeV)^{5}\ s^{-1} 
\earr
\ee
This form in a log-log plot is a straight line and represents the case where the
GT matrix element has a fixed value independent of energy \cite{arnould-07}.
We particularly mention two important waiting point nuclei $^{129}$Ag and
$^{130}$Cd. For $^{129}$Ag our prediction is 0.124\ s compared to the experimental
half-life of 0.046\ s, whereas the earlier prediction of RPA calculations was
0.172\ s. Similarly for $^{130}$Cd our calculation gives the half-life as 0.294\ s
with the experimental value as 0.162\ s and the earlier prediction was 0.195\ s
\cite{Kratz-79}.

After demonstrating that the model for beta strength distribution works well
we use this for predicting half-lives of the very neutron-rich nuclei for which
the experimental data are not available and which are in the r-process path. 
Table 3 gives the predicted half-lives for 25 such nuclei. These are nuclei
considered to estimate the nuclear shell effects at $N=82$ in the r-process
region \cite{Farhan-06}.

Finally Table 4 shows the beta decay rates calculated for two typical nuclei
$^{130}$Ag and $^{126}$Ru. For the first nucleus the half-life is known
 but not the $\log ft$ to specific states, and for the second even the
half-life is not known. The rates are calculated for a large grid of 
temperature and density. The temperatures are taken in the range of 0.5 to
8 $\times$ $10^{9}$ K and the density in the range of $10^{3}$ to $10^{9}$ g/cc.
As the density increases the chemical potential of the electrons goes up
making fewer beta decays energetically possible. The temperature dependence
is seen to be very mild coming from the phase space integral of the 
electrons. It should be mentioned here that the rates are calculated for
transitions only from the ground state of the mother. However for large enough
temperatures some of the odd-odd nuclei having low-lying states may
have contributions from a few excited states also. But for most of the neutron-rich
nuclei of interest, there are no observed excitation spectra and one has
to use a statistical method to take into account the contributions from the excited states of the mother. The departure from a Gaussian shape can also be taken into account
by including effects of higher moments in a parametric form. The so-called `back resonances' of some special excited states with large overlap with the daughter ground state region are also important 
for the higher temperatures and need to be taken into account. The model
also needs to be extended to higher masses. 

After this work was completed we came to know of the paper by Cuenca-Garcia {\it et al.} \cite{Cuenca-07}, wherein results of large-scale shell model calculations of the half-lives of r-process waiting point nuclei at $N=82$ are presented. A detailed comparison of this work with ours will also be included in a future work.






\newpage

\vspace*{-1in}

\begin{center} 

Table 1. Half-lives of nuclei for which $\log ft$ values are available and were used. 

\medskip

\begin{tabular} {rllrcll}
\hline
     &   Mother  &  Daughter  &  Q-value     & No of lowlying  &     $\tau_{1/2}$    &      $\tau_{1/2}$    \\
      &  nucleus   &  nucleus  &                      &  $\log ft$'s taken      &   Expt     &  Calc  \\
  & & & (MeV) & & (sec) & (sec) \\
\hline
 1.  & $^{116}_{45}$Rh$^{\ \ \ }_{71}$             &     $^{116}$Pd         &          8.900   &      3                    &      0.680 &   0.496         \\
 
 2.  & $^{116}_{47}$Ag$^{\ \ \ }_{69}$         &     $^{116}$Cd            &           6.160  &     5                    &     160.8    &  162.8     \\

 3.  & $^{118}_{47}$Ag$^{\ \ \ }_{71}$             &           $^{118}$Cd             &         7.060   &       4                  &       3.76  &  7.61 \\ 

 4.  & $^{119}_{47}$Ag$^{\ \ \ }_{72}$         &        $^{119}$Cd       &          5.350  &      3                   &    2.1   &23.61          \\

 5.  & $^{120}_{47}$Ag$^{\ \ \ }_{73}$         &        $^{120}$Cd       &          8.2    &      4                  &      1.23  &  0.573        \\

 6. & $^{120}_{49}$In$^{\ \ \ }_{71}$         &           $^{120}$Sn        &        5.370    &       3                  &        3.08 &    2.90    \\

 7.  & $^{121}_{47}$Ag$^{\ \ \ }_{74}$           &       $^{121}$Cd         &         6.400  &      4                  &      0.78   &  2.42        \\

 8.  & $^{122}_{47}$Ag$^{\ \ \ }_{75}$           &           $^{122}$Cd     &        9.1    &      4                 &       0.48   &   0.393      \\

 9. & $^{122}_{49}$In$^{\ \ \ }_{73}$           &           $^{122}$Sn            &        6.370  &      3                 &       1.5    & 59.30       \\

 10. & $^{123}_{47}$Ag$^{\ \ \ }_{76}$         &           $^{123}$Cd             &         7.400    &      5                 &      0.309  & 12.69        \\

 11. & $^{123}_{48}$Cd$^{\ \ \ }_{75}$         &           $^{123}$In             &         6.120    &      3                 &      2.10   &  9.46         \\

 12. & $^{124}_{49}$In$^{\ \ \ }_{75}$         &           $^{124}$Cd            &         7.36   &      3                 &      3.17   &  6.60        \\

 13. & $^{125}_{48}$Cd$^{\ \ \ }_{77}$         &           $^{125}$In            &        7.16    &     4                  &      0.65   &  1.32        \\

 14. & $^{125}_{49}$In$^{\ \ \ }_{76}$         &           $^{125}$Sn      &              5.418   &     6                  &      2.36   &  2.93        \\

 15. & $^{126}_{49}$In$^{\ \ \ }_{77}$         &           $^{126}$Sn            &        8.210   &     4                  &      1.60   &  2.45        \\

 16. & $^{127}_{49}$In$^{\ \ \ }_{78}$         &           $^{127}$Sn            &        6.510   &     4                  &      1.09   &  1.23        \\

 17. & $^{128}_{48}$Cd$^{\ \ \ }_{80}$         &           $^{128}$In            &        7.100   &     4                  &      0.34   &  0.248       \\

& & & & & \\
 & & & & & \\

\end{tabular}

\newpage

Table 1 (contd.)

\begin{tabular} {rllrcll}
\hline
     &   Mother  &  Daughter  &  Q-value     & No of lowlying  &     $\tau_{1/2}
$    &      $\tau_{1/2}$    \\
      &  nucleus   &  nucleus  &                      &  $\log ft$'s taken      &
 Expt     &  Calc  \\
 & & & (MeV) & & (sec) & (sec) \\
\hline

 18. & $^{128}_{49}$In$^{\ \ \ }_{79}$         &           $^{128}$Sn            &        8,980   &     3                  &      0.84   &  2.28        \\ 

 19. & $^{129}_{49}$In$^{\ \ \ }_{80}$         &           $^{129}$Sn            &        7.660   &     2                  &      0.61   &  0.664       \\

 20. & $^{130}_{49}$In$^{\ \ \ }_{81}$         &           $^{130}$Sn            &        10.250  &     2                  &      0.32   &  0.265       \\

  21. & $^{131}_{49}$In$^{\ \ \ }_{82}$         &           $^{131}$Sn            &         9.180  &     3                  &      0.282  &  0.273       \\

 22. & $^{132}_{49}$In$^{\ \ \ }_{83}$         &           $^{132}$Sn            &       13.600  &     4                 &     0.201  &  0.166       \\
 
 23. & $^{132}_{51}$Sb$^{\ \ \ }_{81}$         &            $^{132}$Te           &         5.290  &     5                  &    167.4    & 136.6        \\

 24. & $^{133}_{50}$Sn$^{\ \ \ }_{83}$         &            $^{133}$Sb           &         7.830  &     2                  &      1.44   &   1.18       \\
 
 25. & $^{135}_{52}$Te$^{\ \ \ }_{83}$         &            $^{135}$In           &         5.960  &     3                  &     19.0    &  19.2        \\

 26. & $^{136}_{52}$Te$^{\ \ \ }_{84}$         &            $^{136}$I            &         5.070  &     4                  &     17.5    &  26.8        \\

 27. & $^{136}_{53}$I$^{\ \ \ }_{83}$           &            $^{136}$Xe            &         6.930  &     3                  &     83.4    &  91.8        \\

 28. & $^{137}_{53}$I$^{\ \ \ }_{84}$           &            $^{137}$Xe            &         5.880  &     2                  &     24.5    &  50.4        \\

 29. & $^{138}_{53}$I$^{\ \ \ }_{85}$           &            $^{138}$Xe            &         7.820  &     3                  &      6.49   &   9.57       \\

 30. & $^{138}_{55}$Cs$^{\ \ \ }_{83}$          &            $^{138}$Ba            &         5.373  &     3                  &   2004.6    & 506.9        \\

 31. & $^{139}_{54}$Xe$^{\ \ \ }_{85}$          &            $^{139}$Ba            &         5.057  &     6                  &     39.68   &  70.06       \\

 32. & $^{140}_{55}$Cs$^{\ \ \ }_{85}$          &            $^{140}$Ba            &         6.219  &     5                  &     63.7    &  94.8        \\

\hline

\end{tabular}

\end{center}

\newpage

\begin{center}

Table 2. Half-lives of nuclei for which no $\log ft$ values are available 

\vspace*{-0.01in}

\begin{tabular} {rlllll}
\hline
     &   Mother  &  Daughter  &  Q-value     &      $\tau_{1/2}$    &      $\tau_{1/2}$    \\
      &  nucleus   &  nucleus  &                      &   Expt     &  Calc  \\
 & & & (MeV) & (sec) & (sec) \\
\hline

1.  & $^{116}_{43}$Tc$^{\ \ \ }_{73}$         &       $^{116}$Ru          &          11.6                      &      0.09 &  0.111        \\

2.  & $^{116}_{44}$Ru$^{\ \ \ }_{72}$         &       $^{116}$Rh          &           6.3                      &      0.4   &  0.977        \\

3.  & $^{117}_{43}$Tc$^{\ \ \ }_{74}$         &       $^{117}$Ru        &         10.1                       &      0.04   &  0.144          \\

4.  & $^{117}_{44}$Ru$^{\ \ \ }_{73}$           &           $^{117}$Rh   &          8.9                     &      0.3  & 0.217        \\

5.  & $^{117}_{45}$Rh$^{\ \ \ }_{72}$         &       $^{117}$Pd        &          7.000                     &      0.44  &  0.737         \\

6.  & $^{117}_{46}$Pd$^{\ \ \ }_{71}$         &       $^{117}$Ag        &          5.700                     &      4.3   &  2.08          \\

7.  & $^{118}_{45}$Rh$^{\ \ \ }_{73}$         &       $^{118}$Pd        &          10.4                       &      0.3   &  0.194         \\

8.  & $^{119}_{46}$Pd$^{\ \ \ }_{73}$         &       $^{119}$Ag        &           6.500                     &      0.92  &  1.04          \\

9.  & $^{120}_{46}$Pd$^{\ \ \ }_{74}$         &       $^{120}$Ag        &           5.55                      &      0.5   &  1.85          \\

10. & $^{122}_{45}$Rh$^{\ \ \ }_{77}$         &       $^{122}$Pd        &          11.8                       &      0.05  &  0.107         \\

11. & $^{124}_{46}$Pd$^{\ \ \ }_{78}$         &       $^{124}$Ag        &           7,7                       &      0.2   &  0.383         \\

12. & $^{124}_{47}$Ag$^{\ \ \ }_{77}$         &       $^{124}$Cd        &          10.21                      &      0.172 &  0.200         \\

13. & $^{125}_{47}$Ag$^{\ \ \ }_{78}$         &       $^{125}$Cd        &           8.56                      &      0.188 &  0.307         \\

14. & $^{126}_{47}$Ag$^{\ \ \ }_{79}$         &       $^{126}$Cd        &          11.33                      &      0.107 &  0.135         \\

15. & $^{126}_{48}$Cd$^{\ \ \ }_{78}$         &       $^{126}$In        &          5.490                      &      0.506 &  1.86          \\

16. & $^{127}_{47}$Ag$^{\ \ \ }_{80}$         &       $^{127}$Cd        &          9.62                       &      0.79  &  0.170         \\

17. & $^{127}_{48}$Cd$^{\ \ \ }_{79}$         &       $^{127}$In        &          8,470                      &      0.43  &  0.338         \\

18. & $^{128}_{47}$Ag$^{\ \ \ }_{81}$         &       $^{128}$Cd        &         12.5                        &      0.058 &  0.0961        \\

19. & $^{129}_{47}$Ag$^{\ \ \ }_{82}$         &       $^{129}$Cd        &         10.7                        &      0.046 &  0.124         \\ \\

 & & & & & \\
 & & & & & \\

\end{tabular}

\newpage

Table 2 (cont.)

\begin{tabular} {rlllll}
\hline
     &   Mother  &  Daughter  &  Q-value     &      $\tau_{1/2}$    &      $\tau
_{1/2}$    \\
      &  nucleus   &  nucleus  &                      &   Expt     &  Calc  \\
 & & & (MeV) & (sec) & (sec) \\
\hline

20. & $^{129}_{48}$Cd$^{\ \ \ }_{81}$         &       $^{129}$In        &          9.74                       &      0.29  &  0.194         \\

21. & $^{130}_{47}$Ag$^{\ \ \ }_{83}$         &       $^{130}$Cd        &         15.4                        &      0.050 &  0.0627        \\

22. & $^{130}_{48}$Cd$^{\ \ \ }_{82}$         &       $^{130}$In        &          8.29                       &      0.162 &  0.294         \\

23. & $^{131}_{48}$Cd$^{\ \ \ }_{83}$         &       $^{131}$In        &         12.84                       &      0.068 &  0.0923        \\

24. & $^{132}_{48}$Cd$^{\ \ \ }_{84}$         &       $^{132}$In        &         19.19                       &      0.097 &  0.0598        \\

25. & $^{133}_{49}$In$^{\ \ \ }_{84}$         &       $^{133}$Sn        &         13.0                        &      0.180 &  0.0970        \\

26. & $^{134}_{49}$In$^{\ \ \ }_{85}$         &       $^{134}$Sn        &         14.85                       &      0.140 &  0.0767        \\

27. & $^{134}_{50}$Sn$^{\ \ \ }_{84}$         &       $^{134}$Sb        &          6.800                      &      1.04  &  0.664         \\

28. & $^{134}_{51}$Sb$^{\ \ \ }_{83}$         &       $^{134}$Te        &          8.420                      &      0.85  &  0.386         \\

29. & $^{135}_{49}$In$^{\ \ \ }_{86}$         &       $^{135}$Sn         &         13.6                        &      0.092 &  0.0795        \\

30. & $^{135}_{50}$Sn$^{\ \ \ }_{85}$         &       $^{135}$Sb        &          8.9                        &      0.530 &  0.249         \\

31. & $^{135}_{51}$Sb$^{\ \ \ }_{84}$         &       $^{135}$Te        &          8.120                      &      1.71  &  0.403         \\

32. & $^{136}_{50}$Sn$^{\ \ \ }_{86}$         &       $^{136}$Sb        &          8.4                        &      0.25  &  0.279         \\

33. & $^{136}_{51}$Sb$^{\ \ \ }_{85}$         &       $^{136}$Te        &          9.3                        &      0.82  &  0.250         \\

34. & $^{137}_{50}$Sn$^{\ \ \ }_{87}$         &       $^{137}$Sb        &         10.0                        &      0.190 &  0.158         \\

35. & $^{137}_{52}$Te$^{\ \ \ }_{85}$         &       $^{137}$I         &          6.940                      &      2.495 &  0.670         \\

36. & $^{138}_{62}$Te$^{\ \ \ }_{86}$         &       $^{138}$I         &          6.370                      &      1.4   &  0.948         \\

37. & $^{139}_{53}$I$^{\ \ \ }_{86}$          &       $^{139}$Xe        &          6.806                      &      2.29  &  0.691         \\

\hline

\end{tabular}

\end{center}

\newpage

\begin{center}

Table 3. Predicted half-lives for very neutron-rich nuclei (Q values are from \cite{tabisotope-03})

\begin{tabular} {rllll}
\hline
   & Mother & Daughter & Q-value & Predicted \\
   & nucleus & nucleus &   & $\tau_{1/2}$   \\

   &        &      &  (MeV) &  (sec)   \\
\hline
1. & $^{116}_{36}$Kr & $^{116}_{37}$Rb & $21.30$ & 0.0114  \\

2. & $^{118}_{36}$Kr & $^{118}_{37}$Rb & $21.76$ & 0.0100  \\

3. & $^{120}_{36}$Kr & $^{120}_{37}$Rb & $23.55$ & 0.00881  \\

4. & $^{122}_{36}$Kr & $^{122}_{37}$Rb & $25.71$ & 0.00783  \\

5. & $^{118}_{38}$Sr & $^{118}_{39}$Y  & $18.21$ & 0.0167  \\

6. & $^{120}_{38}$Sr & $^{120}_{39}$Y  & $19.75$ & 0.0141  \\

7. & $^{122}_{38}$Sr & $^{122}_{39}$Y  & $21.55$ & 0.0122  \\

8. & $^{124}_{38}$Sr & $^{124}_{39}$Y  & $22.47$ & 0.0108  \\

9. & $^{120}_{40}$Zr & $^{120}_{41}$Nb & $14.86$ & 0.0278  \\

10. & $^{122}_{40}$Zr & $^{122}_{41}$Nb & $16.68$ & 0.0215  \\

11. & $^{124}_{40}$Zr & $^{124}_{41}$Nb & $18.61$ & 0.0175  \\

12. & $^{126}_{40}$Zr & $^{126}_{41}$Nb & $19.56$ & 0.0152  \\ 

13. & $^{122}_{42}$Mo & $^{122}_{43}$Tc & $11.60$ & 0.0585  \\

14. & $^{124}_{42}$Mo & $^{124}_{43}$Tc & $13.61$ & 0.0375  \\

15. & $^{126}_{42}$Mo & $^{126}_{43}$Tc & $12.82$ & 0.0370  \\
\end{tabular}

\newpage

Table 3 (contd.)

\begin{tabular} {rllll}
\hline
   & Mother & Daughter & Q-value & Predicted \\
   & nucleus & nucleus &   & $\tau_{1/2}$   \\

   &        &      &  (MeV) &  (sec)   \\
\hline

16. & $^{128}_{42}$Mo & $^{128}_{43}$Tc & $15.25$ & 0.0249  \\

17. & $^{124}_{44}$Ru & $^{124}_{45}$Rh & $8.60$  & 0.178   \\

18. & $^{126}_{44}$Ru & $^{126}_{45}$Rh & $9.94$  & 0.101   \\

19. & $^{128}_{44}$Ru & $^{128}_{45}$Rh & $11.60$ & 0.0599  \\

20. & $^{130}_{44}$Ru & $^{130}_{45}$Rh & $16.56$ & 0.0306  \\

21. & $^{126}_{46}$Pd & $^{126}_{47}$Ag & $8.27$  & 0.2619  \\

22. & $^{128}_{46}$Pd & $^{128}_{47}$Ag & $10.39$  & 0.1165  \\

23. & $^{130}_{46}$Pd & $^{130}_{47}$Ag & $12.28$ & 0.07009 \\

24. & $^{132}_{46}$Pd & $^{132}_{47}$Ag & $16.56$ & 0.04706 \\

25. & $^{134}_{48}Cd$ & $^{134}_{49}$In & $11.57$ & 0.09227  \\

\hline
 & & & &  \\
 & & & &  \\
\end{tabular}
\medskip

\newpage

  Table 4. Beta decay rates for the nuclei $^{126}$Ru and $^{130}$Ag

\begin{tabular} {rllllll}
\hline
NUCLEUS   &  DENSITY &   &   & TEMP~~T$_{9}$ &   &    \\
          & (g/cm$^3$) & 0.5 & 1.0  &   2.0  &  4.0  &  8.0  \\
\hline \\
          &  10$^{9}$ & 3.56 & 4.29 & 5.12 & 4.79 & 4.22  \\
          &  10$^{8}$  & 6.27 & 6.29 & 5.74 & 4.90 & 4.24  \\
          &  10$^{7}$  & 6.77 & 6.49 & 5.80 & 4.91 & 4.24  \\
$^{126}$Ru &  10$^{6}$  & 6.80 & 6.51 & 5.80 & 4.91 & 4.24  \\
          &  10$^{5}$  & 6.80 & 6.51 & 5.80 & 4.91 & 4.24  \\
          &  10$^{4}$  & 6.80 & 6.51 & 5.80 & 4.91 & 4.24  \\
          &  10$^{3}$  & 6.80 & 6.51 & 5.80 & 4.91 & 4.24  \\ \\
\hline \\
          &  10$^{9}$ & 7.44 & 8.16 & 8.94 & 8.24 & 7.14  \\
          &  10$^{8}$ &10.43 &10.42 & 9.70 & 8.39 & 7.15  \\
          &  10$^{7}$ &10.93 &10.63 & 9.77 & 8.41 & 7.16  \\
$^{130}$Ag &  10$^{6}$ &10.97 &10.65 & 9.78 & 8.41 & 7.16  \\
          &  10$^{5}$ &10.97 &10.65 & 9.78 & 8.41 & 7.16  \\
          &  10$^{4}$ &10.97 &10.65 & 9.78 & 8.41 & 7.16  \\
          &  10$^{3}$ &10.97 &10.65 & 9.78 & 8.41 & 7.16  \\ \\
\hline          
          
\end{tabular}
\medskip

\end{center}

\newpage

\begin{figure}[h]
   \center
   \includegraphics{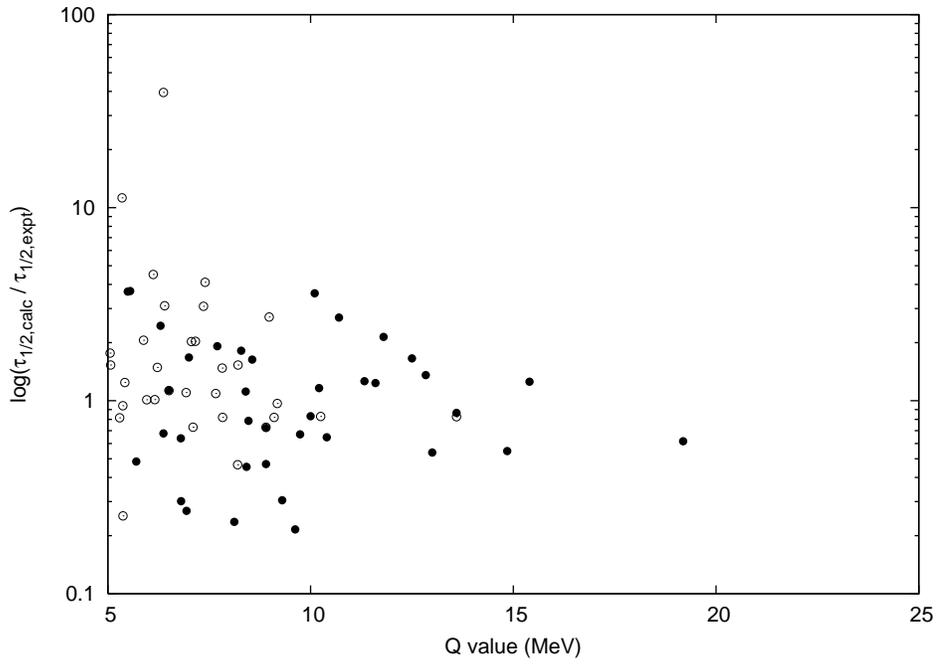}%
    \caption{$\log(\tau_{1/2,\rm calc} / \tau_{1/2,\rm exp})$ plotted against Q value for 69 nuclei (see text). Results for nuclei with known (unknown) log ft's are shown as open (filled) circles. }
\label{}
\end{figure}

\newpage

\begin{figure}[h]
  \center
    \includegraphics{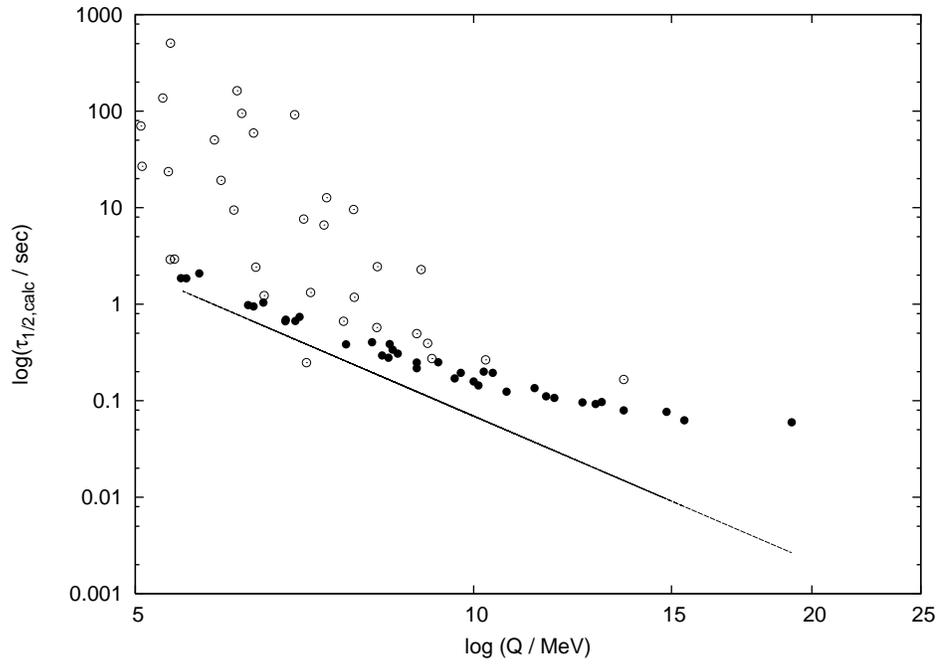}%
    \caption{Calculated half-lives of 69 nuclei plotted against Q values in a log-log plot. Results for nuclei with known (unknown) log ft's are shown as open (filled) circles. A power law dependence is shown for comparison (see text). }
\label{}
\end{figure}

\ed